\documentclass[11pt]{article}

\newif\ifnotes
\notesfalse

\usepackage{hyperref}
\usepackage{fullpage}
\usepackage{amssymb}
\usepackage{amsmath}
\usepackage{amsthm}
\usepackage{amsfonts}
\usepackage{bm}
\usepackage{enumitem}
\usepackage{color}
\usepackage{comment}
\usepackage[capitalize]{cleveref}
\usepackage[dvipsnames]{xcolor}
\usepackage{float}
\usepackage[T1]{fontenc}
\usepackage{todonotes}
\usepackage[most]{tcolorbox}
\usepackage{booktabs}

\usepackage[T1]{fontenc}

\definecolor{DarkRed}{RGB}{150,0,0}
\definecolor{darkturquoise}{rgb}{0.1, 0.6, 0.73}

\usepackage{hyperref}
\hypersetup{
    colorlinks=true,
    linkcolor=darkturquoise,
    filecolor=darkturquoise,
    citecolor=YellowGreen,
    urlcolor=YellowGreen,
}
\usepackage[hyperpageref]{backref}

\newtheorem{theorem}{Theorem}[section]
\newtheorem{definition}[theorem]{Definition}
\newtheorem{lemma}[theorem]{Lemma}
\newtheorem{claim}[theorem]{Claim}

\newtheorem{open}[theorem]{Open Problem}
\newtheorem{proposition}[theorem]{Proposition}

\theoremstyle{remark}

\Crefname{theorem}{Theorem}{Theorems}
\Crefname{claim}{Claim}{Claims}
\Crefname{lemma}{Lemma}{Lemmas}
\Crefname{proposition}{Proposition}{Propositions}
\Crefname{corollary}{Corollary}{Corollaries}
\Crefname{definition}{Definition}{Definitions}

\newcommand{\ECC}{\mathsf{ECC}}
\newcommand{\DEC}{\mathsf{DEC}}
\newcommand{\supp}{\mathsf{supp}}

\makeatletter
\newcommand{\customlabel}[2]{%
   \protected@write \@auxout {}{\string \newlabel {#1}{{#2}{\thepage}{#2}{#1}{}} }%
   \hypertarget{#1}{#2}
}
\makeatother

\newcommand{\protocol}[3]{
    \stepcounter{figure}
    \vspace{0.15cm}
    { \small
    \begin{tcolorbox}[breakable, enhanced, colback=darkturquoise!10]
    \begin{center}
    {\bf \underline{Protocol~\customlabel{prot:#2}{\thefigure}: #1}}
    \end{center}
    
    #3
    \end{tcolorbox}
    }
}

\begin{document}

\title{On Interactive Coding Schemes with Adaptive Termination}
\author{Meghal Gupta \thanks{E-mail:\texttt{meghal.gupta@gmail.com}. This research was supported by a U.C. Berkeley Chancellor's Fellowship.}\\U.C. Berkeley \and Rachel Yun Zhang\thanks{E-mail:\texttt{rachelyz@mit.edu}. This research was supported in part by DARPA under Agreement No. HR00112020023, an NSF grant CNS-2154149, and NSF Graduate Research Fellowship 2141064.}\\MIT}
\date{\today}

\sloppy
\maketitle
In interactive coding, Alice and Bob wish to compute some function $f$ of their individual private inputs $x$ and $y$. They do this by engaging in an interactive protocol to jointly compute $f(x,y)$. The goal is to do this in an error-resilient way, such that even given some fraction of adversarial corruptions to the protocol, both parties still learn $f(x,y)$.

Typically, the error resilient protocols constructed by interactive coding schemes are \emph{non-adaptive}, that is, the length of the protocol as well as the speaker in each round is fixed beforehand. The maximal error resilience obtainable by non-adaptive schemes is now well understood~\cite{BravermanR11,GuptaZ22a,GuptaZ23}. In order to circumvent known barriers and achieve higher error resilience, the work of \cite{AgrawalGS16} introduced to interactive coding the notion of \emph{adaptive} schemes, where the length of the protocol or the speaker order are no longer necessarily fixed.

In this paper, we study the power of \emph{adaptive termination} in the context of the error resilience of interactive coding schemes. In other words, what is the power of schemes where Alice and Bob are allowed to disengage from the protocol early? We study this question in two contexts, both for the task of \emph{message exchange}, where the goal is to learn the other party's input.

\begin{itemize}
\item 
    The first setting is the \emph{full termination} model, where Alice and Bob's order of speaking is fixed but they may terminate at any point. Errors are counted as a fraction of the total number of rounds until the second party has terminated. This is a strengthening of a model proposed by~\cite{AgrawalGS16} that counts errors relative to the last round in which a party speaks: our model disallows the usage of silence to freely communicate information. We construct a protocol achieving $\approx 0.293$ error resilience in this model and show an upper bound of $0.358$. We also demonstrate that in the weaker model of~\cite{AgrawalGS16}, one can utilize silence in a key way to communicate information to obtain a protocol resilient to $\approx 0.3625$ fraction of errors, improving upon~\cite{AgrawalGS16}'s construction achieving $\frac13$ error resilience.
\item 
    We also consider communication over channels with \emph{feedback}. In this setting, Alice and Bob both learn at every point what has been received by both parties. This allows Alice and Bob to use the feedback to agree on a speaking order and agreed termination point for the protocol. In this model, we construct a protocol resilient to $\frac12$ fraction of errors using just a ternary alphabet, which has a natural matching upper bound. In the case of a binary alphabet, we extend the upper bound of $\frac13$ due to~\cite{EfremenkoGH16} for protocols with adaptive speaking order to protocols that have adaptive termination as well. 
\end{itemize}  

\thispagestyle{empty}
\newpage
\tableofcontents
\pagenumbering{roman}
\newpage
\pagenumbering{arabic}

\section{Introduction}

Interactive coding is an interactive analog of error-correcting codes \cite{Shannon48, Hamming50}, that was introduced in the seminal work of Schulman~\cite{Schulman92,Schulman93,Schulman96} and has been an active area of study since.
While error-correcting codes address the problem of sending a {\em message} in a way that is resilient to error, interactive coding addresses the problem of converting an {\em interactive protocol} to an error resilient one. 

In this work, we will be interested in the more specific task of \emph{message exchange}. In message exchange, Alice and Bob each have a private input denoted $x$ and $y$ respectively. The goal is for both parties to learn the other party's input, over the course of an interactive protocol. We will be interested in noise resilient protocols for message exchange, namely, both Alice and Bob should be able to learn the other party's input despite some fraction of the symbols communicated being (adversarially) corrupted. We remark that any interactive protocol can be simulated by message exchange, though not necessarily efficiently.


Usually, interactive coding is studied in the context of \emph{non-adaptive} schemes, which are protocols where the order of speaking and the total length of the protocol are fixed beforehand. It is known that the maximal possible error resilience of any non-adaptive protocol is $\frac14$ if the protocol is over a large alphabet~\cite{BravermanR11} and $\frac16$ if the protocol is over the binary alphabet~\cite{GuptaZ22a,GuptaZ23}. However, in many applications, there's no reason that a protocol has to be a specific length. Indeed, one can often save time by adaptively choosing when to disengage with a protocol, after obtaining the needed information. In this paper, we ask the following question: \emph{What is the power of adaptive termination?} In particular, can the choice of when to terminate alone increase the maximal error resilience?

We study this question in two settings: in the \emph{full termination model}, which is a formalization where two parties speak in a fixed order but may each adaptively choose when to disengage, and in the context of \emph{feedback}, where both parties know at every point every symbol received by either party. In the latter setting, feedback allows the parties to come to a consensus on a speaker each round, and when to terminate, based on a shared transcript of the protocol thus far. This is in contrast to the full termination model, where the views of the two parties may differ so that they cannot sync their termination strategy. 

\subsection{The Full Termination Model}

We first discuss the full termination model. We consider the case where the order of speaking is non-adaptive, that is, in every round there is a pre-specified speaker and listener, so that we can focus on the difficulties introduced by the ability to adaptively terminate. 



We believe that a good model should satisfy the following properties:
\begin{itemize}
    \item {\bf The adversary's budget should be based on the number of symbols sent/received containing information.} Our overarching goal is to design a model where the adversary's budget is determined by the number of meaningful symbols transmitted throughout the protocol.
    \item {\bf No free information.} Silence should not be used to communicate information, or it should count towards the adversary's budget.
\end{itemize}

Since the parties have a fixed order of speaking, each party only needs to make the adaptive decision of when to terminate.

The no free information property ensures that as long as any party has not terminated, each round should count towards the adversary's budget. This is true even if the other party has already terminated. In particular, (1) the rounds where the other party listens should still count towards the budget because they are gaining information that the other party has terminated, and (2) the rounds where the second party speaks still count towards the adversary's budget because they have a chance of communicating meaningful information in the sender's eyes.

Also, as exactly one symbol is sent per round, each round can count equally towards the adversary's budget.

This results in the following model for adaptive termination, which we call the full termination model.

\paragraph{Model:} Alice and Bob follow a fixed order of speaking. In each round, each player may independently make an irreversible decision to terminate. When a party terminates, he/she gives an output and stops participating in the protocol. From that point on, assuming that the other party has not yet terminated, a $\emptyset$ is automatically sent in each round in which the terminated party is the speaker. The adversary may corrupt any round, including rounds in which a $\emptyset$ is sent.
\begin{itemize}
\item {\bf Correctness.}
    We will be interested in the \emph{message exchange protocol}, in which Alice's goal is to tell Bob her input $x$ and Bob's goal is to tell Alice his input $y$. A run of the protocol is \emph{correct} if Alice and Bob both output $(x, y)$.
\item {\bf Adversarial Noise Budget.}
    The noise rate of a rendition of the protocol is the fraction of corrupted rounds out of the total number of rounds until the second party terminates. We will refer to this as \emph{noise rate relative to full (both parties') termination}.
\end{itemize}

\noindent
Our thesis is that the full termination model is the right model for adaptive termination.

\subsubsection{Comparison to The Model of~\cite{AgrawalGS16} (Speaker Termination Model)}

The work of~\cite{AgrawalGS16} defines an adaptive termination model, which we call the \emph{speaker termination} model, in a similar vein to ours: the order of speaking is fixed, and the parties are allowed to terminate at any point, at which point they must output and completely disengage from the protocol. The difference is that the noise budget they allow the adversary is a fraction of the number of rounds that either party speaks, as opposed to of the total number of rounds until both parties have terminated. 
Under the speaker termination model, \cite{AgrawalGS16} constructs a protocol that remains correct even when $\frac13$ of the sent bits are corrupted. We remark that here we are concerned with schemes over a large alphabet.

However, this speaker termination model does not satisfy the no free information clause as we have stated it. 
In particular, we show that it is possible to abuse this early termination and resulting silence to send information about one's input. 
We improve upon this noise resilience and construct a protocol resilient to a noise rate of $\approx 0.3625$ errors in their model. Our protocol crucially uses the free ``silence'' symbols to communicate useful information. 

\begin{theorem} \label{thm:speaker-lower}
    There is a protocol in the speaker termination model that is resilient to a noise rate of $\frac{9-\sqrt{57}}{4}\approx 0.3625$ relative to the time of the last sent message. 
\end{theorem}

Our construction exploits the weakness in the adaptive termination model of~\cite{AgrawalGS16}, that silence can communicate information but not count towards the adversary's budget. We thus prefer our model, in which noise is counted relative to the full termination of the protocol, where silence cannot be used as a mechanism for free information.

\subsubsection{Our Results in the Full Termination Model}

Thus, we move to the full termination model. We show that the adaptivity provided by the full termination model provides us with a benefit over non-adaptive models. Specifically, we show that the full termination model can achieve a higher error resilience than $\frac14$: in particular, there is a scheme that achieves error resilience $\frac{1}{2+\sqrt{2}}\approx0.2929$. We remark that these results are for communication over large alphabets.

\begin{theorem} \label{thm:full-lower}
    For any $\epsilon>0$, there is a protocol in the full termination model that is resilient to a noise rate of $\frac{1}{2+\sqrt{2}}-\epsilon$ relative to the total number of rounds.
\end{theorem}

Furthermore, we establish that the condition in the full termination model is indeed stricter than the condition in the speaker termination model of \cite{AgrawalGS16}. Specifically, we prove that the highest possible error resilience of any protocol in the full termination model is strictly lower than what can be achieved in the speaker termination model, ($\approx 0.3625$). Consequently, the additional stipulation of full termination prevents the parties from attaining error resilience on par with the speaker termination model. This distinction is necessary to prevent the parties from exploiting silence as free information.

\begin{theorem} \label{thm:full-upper}
    No protocol in the full termination model is resilient to a $0.358$ fraction of the total number of rounds.
\end{theorem}

\subsection{Schemes with Feedback}

Next, we pivot our discussion to a different setting of interactive coding, namely interactive communication protocols with \emph{feedback}. 
In a scheme with feedback, the sender of a message learns the value of the message as received by the other party after every round of communication. One exceptional property of such schemes is that the order of speaking need not be fixed beforehand, yet the parties can still agree on a speaker for each round based on their shared knowledge of the messages received by both sides. 

In~\cite{EfremenkoGH16}, they studied the maximal noise resilience of protocols\footnote{This was done in the general interactive coding setting, where the goal is to make \emph{protocols} noise resilient, as opposed to simply message exchange.} with feedback and a non-fixed order of speaking, yet with fixed protocol length. They showed that for binary (and large) alphabets, there exists a protocol in this model that is resilient to a $\frac13 - \epsilon$ fraction of errors (as a fraction of the total protocol length). They also showed that $\frac13$ is an upper bound on the maximal noise resilience, assuming fixed length. 

We observe that in the feedback setting, the parties are not only able to agree on a (non-fixed) speaker for each round but in fact, are able to coordinate a mutual (possibly early) termination based on the transcript. A coordinated early termination was not studied by~\cite{EfremenkoGH16}.

We show that for ternary (and larger) alphabets, allowing early termination circumvents the $\frac13$ upper bound on noise resilience, at least for the task of message exchange. In particular, there exists a message exchange protocol in the feedback model with non-fixed speaking order and termination that is resilient to a $\frac12 - \epsilon$ fraction of corruptions. This is matched by a natural upper bound of $\frac12$. 

\begin{theorem} \label{thm:feedback-prot}
    There is a variable-length protocol for the task of message exchange in the feedback model with a ternary alphabet that is resilient to a $\frac12 - \epsilon$ fraction of errors.
\end{theorem}

We leave open the question of constructing or disproving a protocol resilient to $\frac12$ errors for general protocols, as opposed to simply message exchange. In the binary case, however, we show that non-fixed termination does not improve the error resilience and prove an upper bound of $\frac13$.

\begin{theorem} \label{thm:feedback-upper}
    No variable-length protocol in the feedback model with a binary alphabet is resilient to a $\frac13$ fraction of corruptions.
\end{theorem}

\subsection{Summary of Results}

Below, we summarize our results along with existing state-of-the-art.

\begin{figure}[h!]
\centering
\begin{tabular}{llll}
    \toprule
    \textbf{Model} & \textbf{Alphabet Size} & \textbf{Protocol} & \textbf{Impossibility Bound} \\
    \midrule
    Full Termination & Large & $\frac{1}{2+\sqrt2} \approx 0.2929$ & $\approx 0.358$ \\
    \addlinespace[0.5em]
    Speaker Termination & Large & $\frac{9-\sqrt{57}}{4} \approx 0.3625$ & $\frac12$~\cite{AgrawalGS16} \\
    \addlinespace[0.7em]
    Adaptive with Feedback & Binary & $\frac13$~\cite{EfremenkoGH16} & $\frac13$  \\
    \addlinespace[0.7em]
    Adaptive with Feedback & Ternary \& larger & $\frac12$ & $\frac12$ (folklore) \\
    \bottomrule
\end{tabular}
\caption{Variable length interactive coding schemes}
\label{fig:adaptive-schemes}
\end{figure}

\subsection{Open Problems and Discussion}

Our work raises many interesting open questions. To start, one immediate open question pertaining to the full termination model is the optimal error resilience.

\begin{open}
    Find the largest constant $c$ so that there is a protocol in the full termination model that is resilient to a noise rate of $\approx c$ relative to the total number of rounds.
\end{open}

Secondly, in our discussion of both the full termination model and the adaptive feedback model, the protocol constructions we provide are for message exchange. We ask if our protocol constructions can be translated to the general setting for \emph{any} interactive two-party computation. The challenge here would be to simulate any noiseless protocol in an efficient manner, resulting in a protocol whose maximal length is proportional to the length of the original protocol.

\begin{open}
    Does there exist a constant $c>\frac14$ so that there is a noise resilient protocol $\pi$ for every noiseless protocol $\pi_0$, where $|\pi_0| = O(|\pi|)$, that is resilient to a noise rate of $\approx c$ relative to the total number of rounds in the full termination model?
\end{open}

\begin{open}
    Is there a variable-length protocol $\pi$ for every noiseless protocol $\pi_0$ in the feedback model with a ternary (or larger) alphabet that is resilient to a $\frac12 - \epsilon$ fraction of errors, with the condition that the maximal length of $\pi$ is at most a constant times larger than $\pi_0$?
\end{open}

Finally, we revisit combining the notions of adaptive length and adaptive order of speaking. A multitude of works, for example, \cite{EfremenkoKS21a} discuss protocols with an adaptive order of speaking. We ask what is the most appropriate model that combines both modes of adaptivity in light of our new model of adaptive termination, and what is the optimal resilience of this model.

\subsection{Related Work}

\paragraph{Coding for Interactive Communication.}
Non-adaptive interactive coding (when the protocol is fixed length and fixed speaking order) was studied starting with the seminal works of Schulman~\cite{Schulman92,Schulman93,Schulman96} and continuing in a prolific sequence of followup works, including~\cite{BravermanR11,Braverman12,BrakerskiK12,BrakerskiN13,Haeupler14,BravermanE14,DaniHMSY15,GellesHKRW16, GellesH17,EfremenkoGH16,GhaffariH13,GellesI18,EfremenkoKS20b,GuptaZ22a,GuptaZ23}.

We refer the reader to an excellent survey by Gelles~\cite{Gelles17} for an extensive list of related work. 

\paragraph{Adaptivity in Interactive Coding.}
The work of \cite{AgrawalGS16} introduces the idea of adaptivity into interactive coding, suggesting that Alice and Bob do not need to fix the order of speaking and length of $\pi$ beforehand. This could potentially allow for a larger error resilience than $\frac14$, and in fact, \cite{AgrawalGS16}, followed by~\cite{GhaffariHS14, EfremenkoKS20, EfremenkoKS21a}, show a variety of schemes achieving resilience greater than $\frac14$. In a practical sense, there is not much reason to impose that Alice and Bob should choose their order and length of speaking beforehand, so this is the more natural model to consider.


There are essentially two aspects of adaptivity that the model needs to deal with. These are (1) adaptivity in the order of speaking and (2) adaptivity in the length of the protocol. Both aspects were brought into consideration by \cite{AgrawalGS16}. 

The first aspect is adaptivity in the order of speaking. Here, the speak-or-listen model has generally been accepted, where in each round a party can choose beforehand whether they will speak or listen in a given round. There is a special silence symbol that a party hears if the other party did not speak. The adversary's budget is a fraction of the total number of rounds in the protocol. The work of \cite{EfremenkoKS21a} shows the optimal error resilience of $\frac5{16}$ in this setting. The second type of adaptivity is in the length of the protocol. This was addressed in the speaker termination model~\cite{AgrawalGS16}, which as we discussed has its own shortcomings.


In the feedback setting, the work of \cite{EfremenkoGH16} introduces and analyzes the optimal error resilience of protocols with an adaptive order of speaking. Adaptivity is especially natural in this setting, because the parties have a shared transcript and thus can agree on an order of speaking. The work of \cite{EfremenkoKPS23} discusses adaptivity in the multi-party setting. In this setting, they are able to provide a transformation from adaptive protocols to noise-resilient adaptive protocols, rather than from non-adaptive protocols to noise-resilient adaptive protocols.

With the exception of~\cite{AgrawalGS16}, past work has focused largely on protocols with an adaptive order of speaking, while we focus on protocols with an adaptive order of termination.

\subsection{Outline}

The rest of the paper is organized as follows. In Section~\ref{sec:overview}, we describe the methods used to show each of our results at a high level. In Section~\ref{sec:models}, we formally define the full termination, speaker termination, and adaptive feedback models. In Section~\ref{sec:2term}, we present our protocol construction and impossibility bound for the full termination model. In Section~\ref{sec:aterm}, we present our protocol construction for the speaker termination model. Finally, in Section~\ref{sec:feedback}, we present our protocol constructions for the adaptive feedback model, and discuss the related open questions.

\section{Technical Overview} \label{sec:overview}

In this section, we explain the main ideas behind the results in this paper. Specifically, we outline our protocol and impossibility bound for the full termination model (Theorem~\ref{thm:full-lower} and Theorem~\ref{thm:full-upper} respectively), the protocol for the speaker termination model (Theorem~\ref{thm:speaker-lower}), and the protocol for the adaptive feedback model (Theorem~\ref{thm:feedback-prot}). 

We refer the reader to Section~\ref{sec:models} for a detailed definition of each model. In all the models in this paper, Alice and Bob, both with inputs in $\{0,1\}^n$ wish to communicate their message to the other party (so the other party can communicate some joint function $f(x,y)$) and may use a protocol with a fixed order of speaking but adaptive length and large constant-sized alphabet.

\subsection{Protocol in the Full Termination Model (Theorem~\ref{thm:full-lower})}

In the full termination model, Alice and Bob engage in an interactive protocol with a fixed speaking order, with the caveat that either party may decide to terminate their engagement at any point. The adversary's corruption is measured as a fraction of the total number of rounds until the last party has terminated. 

Our protocol achieves a resilience of $\frac{1}{2+\sqrt{2}} > 0.2928 > \frac14$, where $\frac14$ is the maximal fraction of noise tolerated by a non-adaptive protocol. 

At a high level, the protocol consists of two parts: in the first part, Alice sends a length $N$ encoding of her input $x$, and then in the second part Bob speaks for some number of rounds until he decides to terminate. Both Alice and Bob may decide to terminate early before all of Bob's allotted rounds have passed. Following the framework of~\cite{AgrawalGS16}, Bob speaks for a number of rounds that scales with the number of corruptions he saw in Alice's rounds. More precisely, if Bob saw $t$ corruptions among Alice's $N$ bits, he sends an encoding of $y$ of length $\sqrt{2}N - 2\sqrt{2}t$. The idea is to choose the length of the protocol as to guarantee that either too many corruptions occurred, or both parties must learn the other's input.

The difficulty is in deciding Alice's policy for terminating. Since she doesn't know the length of Bob's encoding, she doesn't know when to stop listening. 
Since the adversary is permitted to corrupt messages into silence symbols, she cannot accurately assess the length of Bob's message. Instead, she only terminates when she can prove that Bob has already terminated, or there would have been too much corruption. More formally, she computes $E$, the minimum amount of corruption possible if Bob has not yet terminated, and terminates if either $E>\frac{N}{\sqrt{2}}$ or she has received $\sqrt{2}N$ symbols (the most Bob could have sent).

We refer to Section~\ref{sec:full-term-prot} for an in-depth analysis of the parameters. We remark that Proposition~\ref{prop:2-round-full} shows that for two-round protocols (that is, every round where Alice speaks occurs before every round where Bob speaks), our construction is optimal.

\subsection{Impossibility Bound in the Full Termination Model (Theorem~\ref{thm:full-upper})}

In this section, we will outline the impossibility bound that no protocol can have resilience more than $0.358$. Specifically, we need to show that the adversary always has an attack when she is permitted $0.358$ fraction of corruption. 

First, recall that Alice and Bob have a fixed speaking order and that their only adaptivity is when they terminate. We'll start by describing an attack that uses $\approx 0.5$ corruption but show that in order for this attack to actually require $\approx 0.5$ (or even closer to $\approx 0.358$) corruption, the order of speaking in the protocol must follow a very specific structure. Then, we'll show that if the order of speaking follows such a specific structure, there are different attacks that work with less corruption.

For the simple attack using $\approx 0.5$ corruption, let's look at the case where Alice's input is known to be $x$ and Bob's input is known to be one of two options, $y_1$ or $y_2$. The adversary can perform the following attack: corrupt no messages from Alice to Bob, and for the messages from Bob to Alice, alternate between sending messages as though Bob had $y_1$ or $y_2$. In each case, this requires corrupting half of Bob's messages until Alice's termination point, which we call $A$. The ``bad case'' here is where Bob has done all (or in general to require more than $\delta$ corruption, Bob has done more than $2\delta$ fraction) of the talking up to this point. We will try to exploit this structure to devise a different attack for the adversary.

Let us (separately) note the following: at any point where it is possible for Bob to terminate, the adversary has an attack corrupting $\delta$ fraction of the messages, where Alice has spoken $\delta$ fraction of the messages until that part. This is because regardless of the input Alice had, the adversary could corrupt the messages to those forcing Bob to terminate at that point, giving him no way to discern Alice's input.

Now, let us combine these ideas. In the attack that we were discussing earlier, to be resilient to $\delta$ corruption, at point $A$, Bob must have spoken at least $2\delta$ of the messages. Now, we divide the argument into cases based on where Bob terminates in the protocol. This point can be different based on whether he has $y_1$ or $y_2$, so let's define $B_1$ to be the point he terminates at given $y_1$ in this attack, and $B_2$ to be the point he terminates at given $y_2$.

If it holds that $B_1 = B_2 = A$ and the protocol is resilient to $\delta$ fraction of error, then Bob must have spoken more than $2\delta$ of the time, but also by the second argument, Alice must have spoken more than $\delta$ of the time. This is a contradiction when $\delta > 1/3$ so no protocol can be resilient to that much error.

It turns out that if $B_1, B_2 \leq A$, the same argument works: after $B_2$ (wlog, let $B_2\geq B_1$), regardless of whether Bob has $y_1$ or $y_2$, he will send silence symbols. As such, the adversary need not corrupt his messages after $B_2$ in either case. At $B_2$, Alice must have spoken more than $\delta$ of the time, but also the corruption required for the second attack is at most half of what Bob has spoken until $B_2$. Then, one of the two attacks requires at most $\frac13$ corruption.

When $B_1, B_2 \geq A$, one can use a similar argument to show that no protocol is resilient to more than $\frac13$ of errors. The difficult case is the one where Bob terminates before $B$ when he has $y_1$ and after when he has $y_2$. In this case, we are not able to achieve a bound of $\frac13$ but rather only $0.358$. 



At a very high level for this case, we introduce another attack for the adversary that works as follows. Previously, we used an attack that corrupted all the messages spoken by Bob until Alice terminated, thus disclosing no information about Bob's input. Now, we find an attack where Bob's messages need only to be corrupted up to a certain point -- afterwards, he will send silence symbols regardless of his input and so the messages do not need to be corrupted. We refer the reader to Section~\ref{sec:full-upper} for a thorough explanation of the argument in this case. 

\subsection{An Improved Protocol in the Speaker Termination Model (Theorem~\ref{thm:speaker-lower})}

The speaker termination model, introduced by~\cite{AgrawalGS16}, counts corruptions as a fraction of the rounds until both parties stop speaking (as opposed to until both parties have also stopped listening and have completely disengaged with the protocol). Our main result is a protocol that achieves $\frac{9-\sqrt{57}}{4} \approx 0.3625$ error resilience in this setting, beating the previous construction of~\cite{AgrawalGS16} achieving a resilience of $\frac13$. Our construction crucially uses silence as a means to communicate information. 

Here we will outline a simpler protocol achieving $\frac{7}{20}$ error resilience. This simpler $\frac7{20}$ protocol will be discussed in length in Section~\ref{sec:7/20}, while the more complicated $\frac{9-\sqrt{57}}{4}$ protocol will be discussed in Section~\ref{sec:9-sqrt57/4}.


In contrast to the scheme of~\cite{AgrawalGS16}, which can be broken into two parts (Alice sends $\ECC(x)$, then Bob sends a variable length encoding of $y$), our protocol is broken into three parts: Alice first sends $\ECC(x)$ for $N$ bits, then Bob sends a length $\frac73N$ encoding of $y$ \emph{that depends on the result of decoding Alice's first message}, and finally Alice responds with a variable length message where the length is based on whether she believes that Bob decoded to $x$ correctly in her first message.


More specifically, Alice first sends Bob $\ECC(x)$ of length $N$ and of distance $\frac12$. Bob decodes this to a guess $\hat{x}$. Then, Bob sends Alice $\ECC_2(\hat{x},y)$ where $\ECC_2$ is a length $\frac73N$ code of relative distance $\approx \frac12$. Alice decodes this to $x'$ and $\hat{y}$.

At this point, both parties have conveyed their input to the other party. Alice always outputs $\hat{y}$. Now, she can fully control the adversary's budget for the protocol -- the length is determined by how long she speaks in the third part. We will have Alice speak for a number of rounds scaling inversely with the number of corruptions she's witnessed in Bob's message, which guarantees that Alice's output $\hat{y}$ is correct as long as the total number of corruptions is below budget. Specifically, if Alice sees $t$ corruptions in Bob's message, she is permitted to send a message up to length $m=\frac{10}{3}N-\frac{20}{7}t$.

Unfortunately, if she employs this strategy, sending Bob $x$ in a code of length $m$, then the adversary may perform the following attack to confuse Bob. The adversary corrupts Alice's messages in the first and third parts to be halfway between what she'd send if she has $x$ or a second input $x'$, and does not corrupt the second message at all. Then, Bob cannot tell if Alice had $x$ or $x'$. Since Alice witnesses $t = 0$ corruptions in Bob's message, Alice's message in the third part has length $m=\frac{10}{3}N$. This means that the total number of corruptions performed by the adversary is $\frac12(N + m) = \frac{13}{6}N$, which is $\frac{13}{40}<\frac{7}{20}$ fraction of the protocol. 

Thus far, Alice has not made use of the fact that in the second round, Bob told her his guess for $\hat{x}$ (which she recorded as $x’$). As explained earlier, as long as she doesn’t send more than $m$ symbols, Alice can fully trust that $x’=\hat{x}$. If this guess is correct (that is, $\hat{x}=x$), Alice simply terminates the protocol immediately \emph{instead of sending $m$ more symbols}. Then, Bob interprets silence symbols to mean that $\hat{x}$ was correct and outputs that guess if he receives many silence symbols. If the adversary tries the same attack as before, where she corrupts half of the first $N$ symbols and then half of the third round of $\frac{10}3N$ symbols, this no longer works because in the case where Alice truly had $\hat{x}$, the protocol’s length is only $\frac{10}{3}N$, so the corruption budget is much smaller than before. Essentially, the silence symbols serve as evidence of the statement ``$\hat{x}$ was correct’’ without the parties paying budget for these messages.

\subsection{The Adaptive Feedback Model (Theorem~\ref{thm:feedback-prot})}

Alice and Bob have inputs $x,y\in \{0,1\}^n$. In the model with feedback, when a party sends a message, they immediately learn what the other party actually received. This means that Alice and Bob have a shared view of the transcript thus far, and can make shared decisions about who speaks each round and when to terminate. Our model captures both these properties.

In this model, we construct a protocol for message exchange using a ternary alphabet that achieves $\frac12$ error resilience. This is optimal: one can imagine an adversary who corrupts half of the bits Alice's sends, corrupting her bits exactly half the time to look like she had a different input, so that Bob cannot determine which of the two inputs she has. 

At a high level, our protocol consists of Alice first sending her input $x$ to Bob over a variable number of rounds, then Bob sending his input to Alice over a variable number of rounds. In order for a party (say Alice) to send her input to Bob, she send symbols $0, 1$, or $\leftarrow$ one at a time to build this guess from scratch. Bob tracks a guess $\hat{x}$ for Alice's input, and at every step, Alice communicates either $0$ or $1$ to indicate that is the next bit of her input, or $\leftarrow$ to indicate that his current guess $\hat{x}$ is an incorrect prefix for her input and he should rewind by $1$. We remark that this ``rewind-if-error'' procedure for building a guess has shown up before, e.g.~\cite{EfremenkoGH16,GuptaZ22a}.

The number of rounds Alice spends to communicate her input is variable. Specifically, she stops communicating when $\hat{x}$ reaches a certain length rather than after a specific number of rounds. At this point, Alice and Bob will switch and now it becomes Bob's turn to communicate $y$ to Alice via the same method.

One can show that in this protocol, the first $\epsilon$ fraction of the bits of $\hat{x}$ and $\hat{y}$ must be correct when the adversary is allocated $\frac12 - \epsilon$ budget for corruption. To ensure that the entirety of $x$ is conveyed correctly, Alice will instead aim to communicate $x$ concatenated with $\approx \frac1\epsilon$ of $0$'s instead of just $x$, so if the first $\epsilon$ of Bob's guess for that is correct, he'll know all of $x$.




We also prove an upper bound of $\frac13$ on the maximal possible error resilience over the \emph{binary} alphabet. This argument follows similar techniques to the upper bound on the error resilience of \emph{non-adaptive} feedback protocols~\cite{EfremenkoGH16}. The main idea is to consider three possible inputs for both Alice and Bob, and have the adversary corrupt each bit to the ``majority'' bit.


\section{Model Definitions} \label{sec:models}

We define the interactive communication models used in this paper. We note that all models, both for lower and upper bounds in this paper are deterministic.

\begin{definition} [Speaker Termination Model]
    A two-party interactive coding scheme in the speaker termination model $\pi$ for a function $f(x, y) : \{0,1\}^n \times \{0,1\}^n \rightarrow \{ 0, 1 \}^o$ is an interactive protocol defined as follows. In each transmission, a single party fixed beforehand sends a single symbol in some alphabet $\Sigma$ to the other party. At any round, a party can opt to terminate and output a guess $\in \{ 0, 1 \}^o$, at which point they can neither speak or listen for the remainder of the protocol.
    
    We say that $\pi$ is resilient to $\alpha$ fraction of adversarial errors if the following holds. For all $x,y \in \{ 0, 1 \}^n$, and for all adversarial attacks where the last party to speak does so at round $R$, if the number of errors is at most $\alpha R$, then Alice and Bob both output $f(x,y)$ at the end of the protocol. 
\end{definition}

\begin{definition} [Full Termination Model]
    A two-party interactive coding scheme in the full termination model $\pi$ for a function $f(x, y) : \{0,1\}^n \times \{0,1\}^n \rightarrow \{0,1\}^o$ is an interactive protocol defined as follows. In each transmission, a single party fixed beforehand sends a single symbol in some alphabet $\Sigma$ to the other party. At any round, a party can opt to terminate and output a guess $\in \{ 0, 1 \}^o$, at which point they can neither speak or listen for the remainder of the protocol.
    
    We say that $\pi$ is resilient to $\alpha$ fraction of adversarial errors if the following holds. For all $x,y \in \{ 0, 1 \}^n$, and for all adversarial attacks where the last party to terminate does so at round $R$, if the number of errors is at most $\alpha R$, then Alice and Bob both output $f(x,y)$ at the end of the protocol. 
\end{definition}

In general, when we discuss the error resilience of schemes in these two models, we view $|\Sigma|$ as large but constant; i.e. $O_\epsilon(1)$ when we are trying to achieve an error resilience of $\alpha-\epsilon$.

\begin{definition} [Adaptive Feedback Model]
    A two-party interactive coding scheme in the full termination model $\pi$ for a function $f(x, y) : \{0,1\}^n \times \{0,1\}^n \rightarrow \{0,1\}^o$ is an interactive protocol defined as follows. In each transmission, a single party sends a single symbol in some alphabet $\Sigma$ to the other party. The party sending the symbol learns what the other party received, allowing the parties to construct a shared transcript $T$ of received bits. In every round, the party who speaks is a function of $T$ (so the parties both agree on the speaker), and whether the parties terminate (which they may do at any time) is a function of $T$.
    
    We say that $\pi$ is resilient to $\alpha$ fraction of adversarial errors if the following holds. For all $x,y \in \{ 0, 1 \}^n$, and for all adversarial attacks where the parties terminate in round $R$, if the number of errors is at most $\alpha R$, then Alice and Bob both output $f(x,y)$ at the end of the protocol. 
\end{definition}
\section{Error Resilience in the Full Termination Model} \label{sec:2term}

\subsection{$1/(2+\sqrt2)$ Construction} \label{sec:full-term-prot}

In this section, we present a protocol resilient to $\frac{1}{2+\sqrt{2}}$ errors in the full termination model. This scheme illustrates that even in the full termination model, the parties can bypass the lower bound of $\frac14$ resilience of non-adaptive schemes.

\protocol{Scheme with Resilience $\frac{1}{2+\sqrt{2}}$ in the Full Termination Model}{1/2rt2}{
        \begin{enumerate}\setcounter{enumi}{-1}
        \item 
            Alice and Bob decide on two codes. $\ECC_N : \{ 0, 1 \}^{n_A} \rightarrow \Sigma^{N}$, with $|\Sigma| = O_\epsilon(1)$ and a relative distance of $1 - \frac12 \epsilon$, and similarly $\ECC_{\sqrt{2}N} : \{ 0, 1 \}^{n_B} \rightarrow \Sigma^{\sqrt{2}N}$.
        \item 
            Alice encodes $x$ with $\ECC_{N}$ of length $N$ and sends the encoding $\hat{x}$ to Bob.
        \item 
            Bob receives word $\hat{x}$ and computes $t = \Delta(\hat{x}, \ECC_N(\DEC(\hat{x}))$. If $t \ge \frac{1-\epsilon}{2} \cdot N$, he terminates. Otherwise, let $L = \sqrt{2}N-2\sqrt{2}t$
            Bob sends Alice the first $L$ letters of $\ECC_{\sqrt{2}N}(y)$. 
        \item 
            After Bob finishes sending $\ECC_L(y)$, he outputs $(\DEC(\hat{x}), y)$ and terminates.
        \item 
            Define $T(x, n)$ to be the first $n$ bits of $x$ if $x$ is longer than $n$, and if shorter, let it be $x$ padded with 0's at the end, so the length is $n$.
            
            When Alice has so far received $R$ bits, call her received string $y_R$. 
            Let $$E_R=\Delta(y_R, T(\ECC_{\sqrt{2}N}(\DEC(T(y_R,\sqrt{2}N))),R)).$$ Alice keeps listening until $E_R> \frac{N}{\sqrt{2}}$ or she has received $\sqrt{2}N$ bits. She outputs $\DEC(T(y_R,\sqrt{2}N))$.
        \end{enumerate}
}

\begin{theorem}
    Protocol~\ref{prot:1/2rt2} is resilient to a $\left(\frac{1}{2+\sqrt{2}}-4\epsilon\right)$-fraction of noise relative to full termination.
\end{theorem}

\begin{proof}
    In this protocol, Alice sends a single (many round) message, Bob sends a single message and terminates, and at some point during or after Bob's message, Alice terminates. 

    Let us first assume that Bob calculates $t\geq \frac{1-\epsilon}{2} \cdot N$ and immediately terminates. Then, the adversary corrupted at least $t$ bits of Alice's message. By default, Alice terminates immediately after round $N+\sqrt{2}N-2\sqrt{2}t+\frac{\sqrt{2}}{2}N$ rounds, and the adversary can only increase this number by one per corruption introduced after Alice's first message. Accordingly, we let the number of corruptions after Alice's first message be $\delta$. Then, the total fraction of corruption by the adversary is at least 
    \begin{align*}
        \frac{t+\delta}{N+\sqrt{2}N-2\sqrt{2}t+\frac{\sqrt{2}}{2}N+\delta} \geq &~ \frac{\frac{1-\epsilon}{2} \cdot N}{N+\sqrt{2}N-(1-\epsilon)\cdot \sqrt{2}N+\frac{\sqrt{2}}{2}N} \\
        \geq &~ \frac{(1-\epsilon)}{2N+\epsilon\sqrt{2}N+\sqrt{2}N} \\
        \geq &~ \frac{N}{2+\sqrt{2}}-\epsilon.
    \end{align*}
    
    Next, let us assume Bob decoded Alice's message incorrectly. Then, the adversary corrupted $(1-\epsilon) \cdot N -t$ symbols. Bob terminates after round $N+\sqrt{2}N-2\sqrt{2}t$. As before, Alice terminates at the latest after $N+\sqrt{2}N-2\sqrt{2}t+\frac{\sqrt{2}}{2}N + \delta$ rounds, where the adversary introduces $\delta$ corruptions after Alice's first message. Then, the total fraction of corruption by the adversary is at least
    \[
        \frac{(1-\epsilon) \cdot N-t+\delta}{N+\sqrt{2}N-2\sqrt{2}t+\frac{\sqrt{2}}{2}N+\delta} \geq \frac{(1-\epsilon) \cdot N-t}{N+\sqrt{2}N-2\sqrt{2}t+\frac{\sqrt{2}}{2}N} \geq \frac{(1-\epsilon)}{2+\sqrt{2}}
    \]
    where the last inequality uses that $t\leq \frac{1-\epsilon}{2}\cdot N$.
    
    Finally, let us assume Bob correctly decodes Alice's message. If Alice terminates before Bob, the adversary must've corrupted at least $\frac{\sqrt{2}}{2}N$ symbols so Alice detects this much "noise" and terminates. The procedure is never longer than $N+\sqrt{2}N$ so this already suffices.
    
    If Alice terminates after Bob, we assume she learns the wrong value of $y$. The adversary's corruptions were as follows: $a$ information symbols were turned into incorrect symbols, $b$ into termination symbols. Also, $c$ termination symbols were turned into information symbols, correct or not. Since Alice decoded incorrectly, by default she would terminate by round $N+\frac{\sqrt{2}}{2}N + \epsilon N$. For every type $a$ or $c$ corruption, the adversary can increase this termination point by $1$. Most type $b$ corruptions do not increase Alice's termination point, except the ones that overlap with her perceived decoding at the step she terminates, of which there are at most $2\epsilon N$. Total corruption is therefore at least $$\frac{t+a+b+c}{N+\frac{\sqrt{2}}{2}N+a+c+3\epsilon N}.$$ Also, for Alice to have decoded wrong, we must have $a+c+2\epsilon N>(\sqrt{2}N-2\sqrt{2}t-a-b)$ where the right-hand side represents the contributors to the wrong decoding and the left-hand side to the correct decoding. This implies that 

    \[
        t\geq \frac12 N - \frac{1}{\sqrt{2}}a - \frac{1}{2\sqrt{2}}b - \frac{1}{2\sqrt{2}}c - \epsilon N.
    \]

    Then, the quantity of corruption is at least

    \[
        \frac{(\frac12-\epsilon) N +(1-\frac{1}{\sqrt{2}})a+(1-\frac{1}{2\sqrt{2}})(b+c)}{N+\frac{\sqrt{2}}{2}N+a+c+3\epsilon N}
    \]

    In order to show this expression is larger than $\frac{1}{2+\sqrt{2}}-4\epsilon$, t suffices to notice that each of the following terms individually is larger than that.

    \[
        \frac{(\frac12-\epsilon) N}{N+\frac{\sqrt{2}}{2}N+3\epsilon N}, \frac{(1-\frac{1}{\sqrt{2}})a}{a}, \frac{(1-\frac{1}{2\sqrt{2}})c}{c} \geq \frac{1}{2+\sqrt{2}}-4\epsilon.
    \]
\end{proof}

The following proposition shows that this noise resilience $\frac{1}{2 + \sqrt2}$ is optimal for protocols where Alice speaks first contiguously, and then Bob speaks.

\begin{proposition} \label{prop:2-round-full}
    No adaptive termination protocol, with the structure that every round where Alice speaks occurs before every round where Bob speaks, is resilient to a $\frac{1}{2+\sqrt2}$-fraction of noise relative to full termination with probability greater than $\frac12$.
\end{proposition}

\begin{proof}
    Let $N$ be the number of rounds where Alice is the speaker, so she speaks for the first $N$ rounds and Bob speaks thereafter. Note that the earliest termination point for either Alice or Bob must be after round $N$, since otherwise Bob has no spoken and Alice cannot know Bob's input. For each possible value $x$ of Alice's input, let $\ell(x)$ be the least number of rounds that Bob speaks after which Alice can terminate (over all possible messages Alice receives). For two inputs $x_0, x_1$, let $\ell_0 = \ell(x_0)$ and $\ell_1 = \ell(x_1)$, and suppose that $\ell_0 \le \ell_1$. Also, let $L(y)$ be the number of rounds that Bob speaks if Alice's first $N$ messages all correspond to if she has input $x = 0$.  Let $L_i = L(y_i)$ for $i \in \{ 0, 1, 2 \}$, and suppose WLOG that $L_0 \le L_1 \le L_2$. 
    
    We consider three attacks:
    \begin{itemize}
        \item 
            First, consider when the adversary corrupts half of Alice's $N$ messages so that Bob receives messages that Alice would have sent with input $x_0$ or $x_1$, each with frequency $N/2$. Since Alice terminates at the earliest at round $N + \ell$, the corruption rate of this attack when Alice's input is $x = x_0$ is $\le \frac{N/2}{N + \ell_0}$, and when $x = x_1$ it is $\le \frac{N/2}{N + \ell_1}$. This attack requires $\le \max ( \frac{N/2}{N + \ell_0}, \frac{N/2}{N + \ell_1} ) = \frac{N/2}{N + \ell_0}$ corruptions.
        \item 
            Now, consider when Alice has input $x_0$, and Bob has either $y_1$ or $y_2$. The adversary corrupts none of Alice's messages so that Bob sends $L_1$ or $L_2$ messages respectively, and she corrupts the first $\ell_0$ of Bob's messages so that Alice terminates after $N + \ell_0$ rounds. Notice that Alice cannot tell which input Bob has. This attack requires $\le \frac{\ell_0}{N + L_i}$ corruptions, which is at most $\frac{\ell_0}{N + L_1}$.
        \item 
            Finally, consider where Alice has $x_0$ and Bob has either $y_0$ or $y_1$, and the adversary corrupts Bob's messages so that Alice receives messages as if Bob had $y_0$ or $y_1$ alternatingly until Alice terminates. If Alice terminates after $\delta \le L_0$ rounds of Bob speaking, this attacks requires $\le \max \left( \frac{\delta/2}{N + L_0}, \frac{\delta/2}{N + L_1} \right) = \frac{\delta/2}{N + L_0} \le \frac{L_0/2}{N + L_0} \le \frac{L_1/2}{N + L_1}$ corruptions. If Alice terminates after $\delta \in (L_0, L_1)$ rounds of Bob speaking, then the attack requires $\max \left( \frac{\delta/2}{N + \delta}, \frac{\delta/2}{N + L_1} \right) = \frac{\delta/2}{N + \delta} \le \frac{L_1/2}{N + L_1}$ corruptions. Finally, if Alice terminates after $\delta \ge L_1$ rounds of Bob speaking, the attack requires $\frac{L_1/2}{N + \delta} \le \frac{L_1/2}{N + L_1}$ corruptions. Thus, this attack will always require $\le \frac{L_1/2}{N + L_1}$ corruptions.
    \end{itemize}
    Combining all three attacks, we see that if the adversary is permitted 
    \[
        r := \min \left( \frac{N/2}{N + \ell_0}, \frac{\ell_0}{N + L_1}, \frac{L_1/2}{N + L_1} \right)
    \]
    corruptions, she has a way to attack the protocol and ensure that at least one party outputs incorrectly with probability at least $\frac12$. We claim that $r \le \frac{1}{2 + \sqrt2}$.
    
    Suppose otherwise, so that $\frac{N/2}{N + \ell_0}, \frac{\ell_0}{N + L_1}, \frac{L_1/2}{N + L_1} > \frac{1}{2+\sqrt2} =: \lambda$. From $\frac{N/2}{N + \ell_0} > \lambda$, we get that $\frac{1/2 - \lambda}{\lambda} \cdot N > \ell_0$. Combining this with $\frac{\ell_0}{N + L_1} > \lambda$, we get that $\frac{1/2 - \lambda - \lambda^2}{\lambda^2} \cdot N > L_1$. Finally, this means that
    \[
        \frac{L_1/2}{N + L_1} < \frac{\frac{1/2 - \lambda - \lambda^2}{\lambda^2} \cdot N / 2}{N + \frac{1/2 - \lambda - \lambda^2}{\lambda^2} \cdot N}
        = \frac{1/2 - \lambda - \lambda^2}{1 - 2\lambda}
        = \lambda,
    \]
    which is a contradiction. Therefore, if the adversary is permitted $\frac{1}{2 + \sqrt2}$ corruptions, at least one of Alice and Bob will output the wrong value with probability at least $\frac12$.
\end{proof}

\subsection{Upper Bound on Resilience of Full Termination Protocols} \label{sec:full-upper}

In this section, we show an upper bound of $0.358$ of error resilience of protocols in the full termination model. We begin with a simpler upper bound of $0.4$ since this still improves upon the bound of $0.5$ while being simpler to understand.

We start by showing a lemma that will be used in the proof of both upper bounds.

\begin{lemma} \label{lem:2term-1-p}
For a protocol to be resilient to a noise rate of $\delta$, at any point where a party $P$ can terminate, party $P$ must have spoken at most $1-\delta$ of the time before that point.
\end{lemma}
\begin{proof}
    Let us say some party $P$ given some input $x$ can terminate at a point $N$. The adversary can send $P$ the bits needed to terminate at that point in the protocol while only corrupting the bits the other party speaks, i.e. only $1-(1-\delta)=\delta$ of the bits regardless of the other party's input. Then, $P$ cannot distinguish between any of the inputs the other party may have had.
\end{proof}

\begin{claim}
No adaptive termination protocol is resilient to a $\frac{2}{5}$-fraction of noise in the full termination model with probability greater than $\frac12$.
\end{claim}

\begin{proof}
    Let Alice have input $x$, and Bob have input $y_1$, $y_2$ or $y_3$. Have the adversary never lie to Bob, and send Alice alternating bits from Bob, as though he had $y_1$, $y_2$ or $y_3$ (so in each case, the adversary would have to corrupt $\frac{2}{3}$ of the bits). Let Alice's termination point receiving these bits be $T$. If Bob has $y_1$, he terminates at $S_1$, if $x_2$, then $S_2$, and if $x_3$ then $S_3$. Two of these are on the same side of $T$; without loss of generality let us suppose these are the options where Bob has $x_1$ and Bob has $x_2$. We show that the proposed attack does not allow Alice to distinguish between Bob having $x_1$ and $x_2$.
    
    If both $S_0$ and $S_1$ are below $T$: At point $S=\max(S_0, S_1)$, Bob has spoken at most $0.6$ of the time by Lemma~\ref{lem:2term-1-p}. Therefore, at most $\frac{2}{3}0.6S=0.4S$ bits are corrupted up to this point with the adversary's attack. No bits need to be corrupted after this point. Because $T > S$, Alice hasn't terminated at this point, so the total rate is at most $0.4S/S = 0.4$.
    
    If both $S_0$ and $S_1$ are above $T$: At point $S=\min(S_0, S_1)$, Bob has spoken at most $0.6$ of the time by the lemma. The adversary's attack only corrupts Bob's bits, and therefore only corrupts at most $\frac{2}{3}0.6S=0.4S$ bits. The adversary never needs to corrupt between $S_0$ and $S_1$ since Alice is guaranteed not listening anymore, so the adversary corrupts at most $0.4S$ bits total. The length of the protocol is at least $S$ in either case, so the adversary stayed within budget.
\end{proof}

\begin{theorem} \label{thm:2term-upper}
    No adaptive termination protocol is resilient to a $0.358$ fraction of noise in the full termination model.
\end{theorem}

\begin{lemma} \label{lem:upper1}
    Let Alice have any possible input $x$ and Bob have any two possible inputs $y_1,y_2$. Given any $\frac13<\delta<\frac12$, in any adaptive termination protocol resilient to $\delta$ fraction of errors, it must be possible for Alice to terminate at a point $A$ where Bob has spoken at least $2\delta$ fraction of the time and for Bob to terminate before $A$ if he has one of $y_1$ or $y_2$ and after $A$ if he has the other.
\end{lemma}

\begin{proof}
    Let Alice have input $x$ and Bob have two possible inputs $y_1, y_2$. The adversary corrupts none of Alice's messages and alternates Bob's messages as though he has $y_1$ and $y_2$, corrupting half the symbols in each case (sending odd-numbered messages as though he has $y_1$ and even-numbered as though he has $y_2$). At the point where Alice terminates, Bob must have spoken at least $2\delta$ fraction of the time otherwise this process resulted in only $\delta$ corruption regardless of which input Bob had. 
    
    If in both cases $y_1$ and $y_2$, Bob terminates before Alice, by Lemma~\ref{lem:2term-1-p} the later of the two points, which we call $B$, falls at a location where Bob has spoken at most $1-\delta$ of the time. Then, between $B$ and $A$, the adversary does not actually need to make any corruptions to alternate Bob's messages to alternate as though he has $y_1$ and $y_2$, since either way he is sending only the termination symbol. Then, the adversary corrupts at most $\frac12\cdot (1-\delta)<\delta$ fraction of the protocol in either case.

    If in both cases $y_1$ and $y_2$, Bob terminates after Alice, we can use a similar argument. By Lemma~\ref{lem:2term-1-p} the earlier of the two points, which we call $B$, falls at a location where Bob has spoken at most $1-\delta$ of the time. Then, to alternate Bob's messages up to the point $B$, the adversary corrupts at most $\frac12\cdot (1-\delta)<\delta$ fraction of the protocol in either case.

    Therefore, in one case Bob must terminate before Alice and in the other case Bob must terminate after Alice.
\end{proof}

\begin{proof} [Proof of Theorem~\ref{thm:2term-upper}.]
    Let Alice and Bob have either inputs $(x_1,\{y_1,y_2\})$, $(x_2,\{y_3,y_4\})$, $(x_3,\{y_5,y_6\})$, $(\{x_4,x_5\},y_7)$, $(\{x_6,x_7\},y_8)$, $(\{x_8,x_9\},y_9)$. In the first three cases, let $A_1,A_2,A_3$ be Alice's termination points from Lemma~\ref{lem:upper1}, and in the last three cases let Bob's termination points from Lemma~\ref{lem:upper1} be $B_1,B_2,B_3$. One of these values is smaller than two values from the other set; i.e. without loss of generality, $A_1<B_1<B_2$. 
    
    At $A_1$, Alice with input $x_1$ has spoken at most $1-2\delta$ fraction of the time, and by Lemma~\ref{lem:upper1} there is a point $B'<A_1$ where Bob can terminate given a specific one of $y_1$ or $y_2$ where Alice has spoken at least $\delta$ fraction of the time. 
    
    Also, Bob with $y_7$ has spoken at most $1-2\delta$ by point $B_1$, so $B_1>\frac{2\delta}{1-2\delta}A_1$. 

    Finally, let $A'>B_1$ be the point at which Alice can terminate given one of $x_4$ or $x_5$ that falls after $B_1$ by Lemma~\ref{lem:upper1}. At $A'$, it holds that Alice has spoken at most $1-\delta$ fraction of the time. Then, it holds that $A'>\frac{1-\delta}{2\delta}B_1$. Similarly, define $A''>B_2$, and since $B_1<B_2$, the same argument asserts that $A''>\frac{1-\delta}{2\delta}B_1$.
    
    Getting Bob with input $y_1$ or $y_2$ to terminate at $B'$ requires at most $(1-2\delta)A_1$ corruption, and getting Alice to terminate at $A'$ or $A''$ (depending on what input she has $x_4$ or $x_5$ versus $x_6$ or $x_7$) requires $(1-2\delta)B_1$ corruption. The protocol length is at least $\min(A',A'')>\frac{2\delta}{1-\delta}B_1$. In total, the fraction of corruption is at most 
    \begin{align*}
        &~\frac{(1-2\delta)A_1+(1-2\delta)B_1}{\frac{2\delta}{1-\delta}B_1} \\
        \leq &~\frac{(1-2\delta)\cdot \frac{1-2\delta}{2\delta}B_1+(1-2\delta)B_1}{\frac{2\delta}{1-\delta}B_1} \\
        = &~\frac{(1-2\delta)\cdot \frac{1-2\delta}{2\delta}+(1-2\delta)}{\frac{2\delta}{1-\delta}} \\
        = &~ \frac{1 - 3\delta + 2\delta^2}{4 \delta^2}
    \end{align*}

    The amount of corruption that Eve must use (assuming $\delta>\frac13$ so the assumptions for Lemma~\ref{lem:upper1} held) is 
    \[
        \min\left(\delta, \frac{1 - 3\delta + 2\delta^2}{4 \delta^2} \right)\leq 0.358
    \]
\end{proof}

\section{Error Resilience in the Speaker Termination Model} \label{sec:aterm}

\subsection{$7/20$ Construction} \label{sec:7/20}

In this section, we present two protocols that achieve resilience higher than the $\frac13$ construction shown by \cite{AgrawalGS16} in the Speaker Termination Model. The second protocol construction achieves a higher resilience than the first, but we present the first as well for its simplicity.

\protocol{Scheme with resilience $7/20 - \epsilon$ in the Speaker Termination Model}{7/20}{
        
        Let $L = \frac73 N$ and $M = \frac{10}3 N$. Alice and Bob decide on a code $\ECC_B : \{ 0, 1 \}^{n_A + n_B} \rightarrow \Sigma^L$ with relative distance $1-\epsilon$, with decoding algorithm $\DEC_B$. Let $\ECC^* : \{ 0, 1 \}^{n_A} \rightarrow \Sigma^{N + M}$ be an error-correcting code with distance $1-\epsilon$ with the property that restricting its output to the first $J$ characters is also an error-correcting code with distance $1-\epsilon$. Let $\ECC_A : \{ 0, 1 \}^{n_A} \rightarrow \Sigma^N$ be the restriction of $\ECC^*$'s output to the first $N$ characters, and let $\DEC_A$ be the corresponding decoding algorithm. For $K \in [M]$, let $\ECC_K : \{ 0, 1 \}^{n_A}$ be the restriction of its output to the slice consisting of all indices between $N+1$ and $N+K$, i.e. $\ECC_K = \ECC^*[N+1,N+K]$. 
        
        \vspace{0.1in}
        On Alice input $x$ and Bob input $y$, the parties do the following:
        \begin{enumerate}
            \item Alice sends Bob $\ECC_A(x)$. 
            \item Bob receives word $m_A \in \Sigma^N$ and decodes $\hat{x}_1 = \DEC_A(m_A)$. He also computes $t = \Delta(m_A, \ECC_A(\hat{x}_1))$. He sends Alice $\ECC_B(\hat{x}_1, y)$.
            \item Alice receives word $m_B \in \Sigma^L$ and decodes $(x', \hat{y}) = \DEC_B(m_B)$. She also computes $s = \Delta(m_B, \ECC_B(x', \hat{y}))$. If $x' = x$ or $s \ge \frac23N$, she terminates immediately. Otherwise, she sets $K = \frac{10}3 N - \frac{20}7 s = M - \frac{20}{7} s$ and sends $\ECC_K(x)$. She then terminates and outputs $\hat{y}$.
            \item Bob listens for $M$ rounds and obtains word $m'_A \in (\{ \perp \} \cup \Sigma)^M$. He computes $\delta_+ = |\supp(m'_A)| = \# \{ i \in [M] : m'_{A,i} \not= \perp \}$. If $\delta_+ < \frac76 N - t$, he outputs $\hat{x}_1$. Otherwise, he computes $\hat{x}_2 = \arg\min_{\hat{x} \not= \hat{x}_1} \Delta(m_A || m'_A, \ECC^*(\hat{x}))$ and outputs $\hat{x}_2$.
        \end{enumerate}

}

\paragraph{Explanation.}
In summary, there are four steps in this protocol.
\begin{enumerate}
    \item Alice sends Bob an encoding of $x$ of a fixed length $N$.
    \item \label{step:2} Bob decodes Alice's message to $\hat{x}_1$ and sends Alice an encoding of $y$ and $\hat{x}_1$ of a fixed length $\frac73N$.
    \item Alice stops immediately if it appears that Bob has already learned $x$ or she can prove there was a lot of corruption. If not, she sends an encoding of $x$ again, with a variable length capped at $\frac{10}{3}N$. (She outputs the guess for $y$ closest to Bob's message.)
    \item \label{step:final} Bob always listens for $M$ rounds since listening is costless. If he witnesses a lot of termination symbols (meaning Alice's message was short or nonexistent), he outputs his original guess $\hat{x}_1$. Otherwise, he combines the encoding he received in Step~\ref{step:2} with the new one, and outputs the decoding of this combined message.
\end{enumerate}

We argue that in Step~\ref{step:final}, Bob is making use of the termination symbols for free information. Alice sends him a new message (rather than termination symbols) only if she believed he previously decoded incorrectly. As such, termination symbols from her end emphatically say ``your previous decoding was correct; stick with it.'' They aren't just telling Bob to terminate, they are actively telling him that his previous decoding $\hat{x}_1$ was correct, and as such, are acting as evidence towards $\hat{x}_1$. Indeed, Bob makes explicit use of this when he outputs his original guess $\hat{x}_1$ upon hearing a lot of termination symbols. These termination symbols are essentially Alice sending the message: ``$\hat{x}_1$ was correct,'' without paying any corruption budget to do so.

In the full termination model, Alice and Bob must pay corruption budget to the adversary for Bob to hear these messages.

\begin{theorem}
    Protocol~\ref{prot:7/20} is resilienct to a $7/20 - \epsilon$ fraction of errors relative to the total number of messages sent. 
\end{theorem}

\begin{proof}
    We split the possible attacks that the adversary can make in three categories and show that in each, the adversary must corrupt at least $\frac7{20} - \epsilon$ of the messages sent.
    
    First, suppose the adversary corrupts messages so that Alice outputs incorrectly. This means that the adversary corrupted $\ge \frac{(1-\epsilon)}2 \cdot L$ of Bob's messages to Alice. Recall that $s = \Delta(m_B, \ECC_B(x', \hat{y}))$, where $(x', \hat{y}) \not= (\hat{x}, y)$. If $s \ge \frac{1-\epsilon}2 \cdot L$ (in fact, if $s \ge \frac23 N$), then Alice immediately terminates, so the total corruption rate is at least $\frac{\frac{1-\epsilon}2 \cdot L}{N + L} = \frac{(1-\epsilon) \cdot \frac76 N}{\frac{10}{3} N} = \frac{7}{20} - \frac{7}{20}\epsilon$. Otherwise, $s < \frac{1-\epsilon}{2} \cdot L$, and the adversary must've corrupted at least $(1-\epsilon)\cdot L-s$ messages. Alice then speaks $\le \frac{10}3 N - \frac{20}7 s$ more times, so that the total number of messages sent in the protocol is $\le N + L + \frac{10}3 N - \frac{20}7 s$. Then, the adversary must've corrupted a fraction of 
    \[
        \ge \frac{(1-\epsilon)\cdot L-s}{N + L + \frac{10}3 N - \frac{20}7 s} = \frac{\frac73(1-\epsilon)\cdot N - s}{\frac{20}3 N - \frac{20}7 s} > \frac{\frac73(1-\epsilon)\cdot N - \frac76 N}{\frac{20}3 N - \frac{20}7 \cdot \frac76 N} = \frac{7}{20} - \frac{7}{10}\epsilon
    \]
    messages, using the fact that $s \le \frac{1-\epsilon}{2} \cdot L < \frac76 N$.
    
    Now, suppose that the adversary corrupts messages so that Bob outputs incorrectly. We also assume that the adversary corrupts fewer than $\frac23 N$ of Bob's messages to Alice: First, we consider the case that the adversary corrupts $k \ge \frac{1-\epsilon}2 \cdot L$ of Bob's messages. If $s \ge \frac{1-\epsilon}2 \cdot L$, then there must've been at least $s$ corruptions, and Alice terminates immediately, giving an error rate of $\ge \frac{\frac{1-\epsilon}{2} \cdot L}{N + L} \ge \frac{7}{20} - \frac{7}{20}\epsilon$. Otherwise, $s \in [(1-\epsilon) \cdot L-k, \frac{1-\epsilon}2 \cdot L)$, with $s \ge (1-\epsilon) \cdot L - k$, and Alice additionally speaks $\le \frac{10}{3} N - \frac{20}{7}s \le \frac{10}{3} N - \frac{20}{7} (1-\epsilon) \cdot \frac73 N + \frac{20}{7}k = \frac{20}{7} k - \frac{10 - 20\epsilon}{3} \cdot N$ more rounds, for a total of $N + L + \frac{20}{7} k - \frac{10 - 20\epsilon}{3} \cdot N = \frac{20\epsilon}{3} N + \frac{20}{7}k$ extra rounds. Then, the adversary must've corrupted a fraction of
    \[
        \ge \frac{k}{\frac{20}{7} k + \frac{20\epsilon}{3} N}
        \ge \frac{\frac{1-\epsilon}2 \cdot L}{\frac{20}{7} \cdot \frac{1-\epsilon}{2} \cdot L + \frac{20\epsilon}{3} N}
        = \frac{\frac{7}6 (1-\epsilon) \cdot N}{\frac{10 + 10\epsilon}{3} N}
        \ge \frac{7}{20}(1-2\epsilon)
        = \frac{7}{20} - \frac{7}{10}\epsilon
    \]
    messages. Otherwise, in the case $k \in [\frac23N, \frac{1-\epsilon}2 L)$ of Bob's messages are corrupted, Alice terminates immediately and has learned Bob's input correctly. Bob will also learn Alice's correct input unless at least $\frac{1-\epsilon}{2} N$ of Alice's messages are corrupted, including messages in the last chunk where Alice does not actually speak. This gives a rate of 
    \[
        \ge \frac{k + \frac{1-\epsilon}2 N}{N + L}
        \ge \frac{\frac23 N + \frac{1-\epsilon}2 N}{\frac{10}{3} N}
        = \frac{7}{20} - \frac{3}{20} \epsilon.
    \]
    
    Now, assuming that fewer than $\frac23N$ of Bob's messages are corrupted, there are three cases where Bob outputs incorrectly. The first is that $\hat{x}_1 = x$ but Bob outputs $(\hat{x}_2, y)$. In this case, since we assumed that fewer than $\frac23 N < \frac{1-\epsilon}2 \cdot L$ of Bob's messages are corrupted, then Alice terminates immediately after Bob finishes speaking since she learns $x' = \hat{x}_1 = x$, so that the total number of messages sent is $N + L = \frac{10}3 N$. But, in order for Bob to output $\hat{x}_2 \not= \hat{x}_1$, he must've heard at least $\frac76 N - t$ non-$\perp$ characters in Alice's last set of messages, for a total of $t + \frac76 N - t = \frac76 N$ corruptions. This gives a corruption rate of $\ge \frac7{20}$.
    
    The second case is that $\hat{x}_1 \not= x$ but Bob outputs $\hat{x}_1$. This can only happen if Bob hears less than $\frac76 N - t$ non-$\perp$ messages in Alice's second chunk. Suppose that the adversary corrupts $s < \frac23 N$ of Bob's messages, so that Alice speaks $\frac{10}3 N - \frac{20}7 s$ additional rounds. Since Bob hears less than $\frac76 N - t$ non-$\perp$ messages, the adversary must've corrupted $> \frac{10}3 N - \frac{20}7 s - (\frac76 N - t) = \frac{13}{6} N - \frac{20}7 s + t$ of Alice's second chunk of messages. Combining with the $\ge (1-\epsilon) \cdot N - t$ corruptions the adversary must've made in the first chunk of Alice's messages, and the $s$ corruptions to Bob's messages, this is a total corruption count of $> ((1-\epsilon) N-t) + s + (\frac{13}{6} N - \frac{20}{7} s + t) = (\frac{19}{6} - \epsilon) \cdot N - \frac{13}{7} s$, giving a corruption rate of 
    \[
        > \frac{(\frac{19}{6} - \epsilon)\cdot N - \frac{13}{7}s}{N + L + \frac{10}{3} N - \frac{20}{7}s}
        = \frac{(\frac{19}{6} - \epsilon) \cdot N - \frac{13}{7}s}{\frac{20}{3} N - \frac{20}{7}s}
        \ge \frac{(\frac{19}{6} - \epsilon) \cdot N - \frac{13}{7} \cdot \frac23 N}{\frac{20}{3} N - \frac{20}{7} \cdot \frac23 N}
        = \frac{81}{200} - \frac{21}{100} \epsilon
        > \frac{7}{20}.
    \]
    
    The third case is that $\hat{x}_1 \not= x$ and Bob outputs $\hat{x}_2 \not= x, \hat{x}_1$. Recall that $t = \Delta(m_A, \ECC_A(\hat{x}_1))$.
    If $t \le \frac23 N$, then the adversary must've corrupted $\ge (1-\epsilon) \cdot N-t$ messages in common between $m_A$ and $\ECC_A(\hat{x}_1)$ as well as at least $(\frac12 t - \epsilon N) + \frac{1-\epsilon}2 \cdot (\frac{10}{3} N - \frac{20}{7} s)$ of Alice's remaining messages (both later messages and ones that are not consistent with $\hat{x}_1$ in the first chunk). Then, the adversary made at least
    \begin{align*}
        \frac{s + ((1-\epsilon)\cdot N-t) + (\frac12 t - \epsilon N) + \frac{1-\epsilon}2 \cdot (\frac{10}{3} N - \frac{20}{7}s)}{N + L + \frac{10}{3} N - \frac{20}{7}s}
        &= \frac{(\frac{8}3 - \frac{11}{3} \epsilon) \cdot N - \frac12 \cdot t - (\frac37 - \frac{10}{7} \epsilon) \cdot s}{\frac{20}{3} N - \frac{20}{7}s} \\
        &\ge \frac{(\frac83 - \frac{11}3 \epsilon) \cdot N - \frac12 \cdot \frac23 N - (\frac37 - \frac{10}{7}\epsilon) \cdot s}{\frac{20}{3}N - \frac{20}{7} s} \\
        &= \frac{(\frac73 - \frac{11}{3} \epsilon) \cdot N - (\frac37 - \frac{10}{7} \epsilon) \cdot s}{\frac{20}{3} N - \frac{20}{7} s} \\
        &\ge \frac{7}{20} - \frac{11}{20} \epsilon
    \end{align*}
    fraction of corruptions.
    
    Alternatively, if $t > \frac23 N$, then in particular $t = \Delta(m_A, \ECC_A(\hat{x}_1)) \le \Delta(m_A, \ECC_A(x)) =: t'$, so there must've been at least $t' \ge t$ corruptions to Alice's first $N$ messages. Furthermore, at least $h(t,t') = \frac{1-\epsilon}2 \cdot (\frac{10}{3} N - \frac{20}{7} s) - \frac{t'-t}{2}$ of Alice's later messages must have been corrupted, otherwise more than $(\frac{10}{3} N - \frac{20}{7} s) - h(t,t') + (N - t')$ of Alice's total messages are consistent with $x$, which is more than the $< \epsilon \cdot (\frac{10}{3} N - \frac{20}{7} s) + h(t, t') + (N - t)$ messages that are consistent with $\hat{x}_2$. This gives a total corruption rate of 
    \begin{align*}
        \frac{t' + s + \frac{1-\epsilon}2 \cdot (\frac{10}{3} N - \frac{20}{7} s) - \frac{t'-t}{2}}{N + L + \frac{10}{3} N - \frac{20}{7} s}
        &\ge \frac{t + (\frac53 - \frac53 \epsilon) \cdot N - (\frac37 - \frac{10}{7}\epsilon) \cdot s}{\frac{20}{3} N - \frac{20}{7} s} \\
        &> \frac{(\frac73 - \frac53 \epsilon) \cdot N - (\frac37 - \frac{10}{7} \epsilon) \cdot s}{\frac{20}{3} N - \frac{20}{7} s} \\
        &= \frac{7}{20} - \frac14 \epsilon.
    \end{align*}
\end{proof}

\subsection{$\frac{9-\sqrt{57}}4$ Construction} \label{sec:9-sqrt57/4}

\protocol{Scheme with resilience $\frac{9-\sqrt{57}}{4} - \epsilon$ in the Speaker Termination Model}{gt1/3}{
        Let $L = \frac{3+\sqrt{57}}4 N$ and $M = \frac{7+\sqrt{57}}{4} N$. Alice and Bob decide on a code $\ECC_B : \{ 0, 1 \}^{n_A + n_B} \rightarrow \Sigma^L$ with relative distance $1-\epsilon$, with decoding algorithm $\DEC_B$. Let $\ECC^* : \{ 0, 1 \}^{n_A} \rightarrow \Sigma^{N + M}$ be an error-correcting code with distance $1-\epsilon$ with the property that restricting its output to the first $J$ characters is also an error-correcting code with distance $1-\epsilon$. Let $\ECC_A : \{ 0, 1 \}^{n_A} \rightarrow \Sigma^N$ be the restriction of $\ECC^*$'s output to the first $N$ characters, and let $\DEC_A$ be the corresponding decoding algorithm. For $K \in [M]$, let $\ECC_K : \{ 0, 1 \}^{n_A}$ be the restriction of its output to the slice consisting of all indices between $N+1$ and $N+K$, i.e. $\ECC_K = \ECC^*[N+1,N+K]$. 
        
        \vspace{0.1in}
        On Alice input $x$ and Bob input $y$, the parties do the following:
        \begin{enumerate}
            \item Alice sends Bob $\ECC_A(x)$. 
            \item Bob receives word $m_A \in \Sigma^N$ and decodes $\hat{x}_1 = \DEC_A(m_A)$. He also computes $t = \Delta(m_A, \ECC_A(\hat{x}_1))$. If $t \ge \frac{9-\sqrt{57}}{4} N$, he terminates and outputs arbitrarily. Otherwise, he sends Alice $\ECC_B(\hat{x}_1, y)$.
            \item Alice receives word $m_B \in \Sigma^L$ and decodes $(x', \hat{y}) = \DEC_B(m_B)$. 
            She also computes $s = \Delta(m_B, \ECC_B(x', \hat{y}))$. If $x' = x$ or $s \ge \frac{\sqrt{57}-1}{8}N$, she terminates immediately. Otherwise, she sets $K(s) = \frac{7+\sqrt{57}}{4} N - \frac{9+\sqrt{57}}{6} s = M - \frac{4}{9-\sqrt{57}} s$ and sends $\ECC_{K(s)}(x)$. She then terminates and outputs $(x, \hat{y})$.
            \item Bob listens for $M$ rounds and obtains word $m'_A \in (\{ \perp \} \cup \Sigma)^M$. He computes $\delta_A = |\supp(m'_A)| = \# \{ i \in [M] : m'_{A,i} \not= \perp \}$. If $\delta_A < \frac{3+\sqrt{57}}{8} N - t$, he outputs $\hat{x}_1$. Otherwise, he computes $\hat{x}_2 = \arg\min_{\hat{x} \not= \hat{x}_1} \Delta(m_A || m'_A, \ECC^*(\hat{x}))$ and outputs $\hat{x}_2$.
        \end{enumerate}
}

\begin{theorem}
    Protocol~\ref{prot:gt1/3} is resilient to a $\frac{9-\sqrt{57}}{4} - \epsilon \approx 0.3625 - \epsilon$ fraction of errors relative to the total number of messages sent. 
\end{theorem}

\begin{proof}
    We proceed by cases on the adversary's succeeding attack.
    
    First, consider any attack in which $t \ge \frac{9-\sqrt{57}}{4} N$. Bob immediately terminates and does not speak. If at most $L - \frac{\sqrt{57}-1}{8} N = \frac{7+\sqrt{57}}{8}$ of Bob's messages are corrupted to be non-$\perp$ characters, then Alice also terminates after Bob's $L$ rounds, so the total number of rounds in which someone speaks is $N$. This gives an error rate of at least $\frac{t}{N} = \frac{9-\sqrt{57}}{4}$. Otherwise, at least $\frac{7+\sqrt{57}}{8} N$ of Bob's messages are corrupted, and Alice speaks at most $M$ more rounds, so that the total number of rounds in which a party speaks is $N + M$. This gives a corruption rate of at least 
    \[
        \frac{t + \frac{7+\sqrt{57}}{8} N}{N + M}
        \ge \frac{\frac{9-\sqrt{57}}{4} N + \frac{7+\sqrt{57}}{8} N}{\frac{11 + \sqrt{57}}{4} N} \approx 0.47 > \frac{9-\sqrt{57}}{4}.
    \]
    Now, for the rest of this analysis, we assume that $t < \frac{9 - \sqrt{57}}{4} N$.
    
    Consider if the adversary corrupts messages so that Alice outputs incorrectly. This means that at least $\frac{1-\epsilon}{2}$ of Bob's messages to Alice are corrupted. In this case, if Alice sees $s$ errors, then there were at least $(1-\epsilon) \cdot L - s$ corruptions and Alice speaks at most $M - \frac{4}{9-\sqrt{57}} s$ more messages. Then, the total corruption rate is at least
    \[
        \frac{(1-\epsilon) \cdot L - s}{N + L + M - \frac{4}{9-\sqrt{57}} s}
        = \frac{(1-\epsilon) \cdot L - s}{\frac{4}{9-\sqrt{57}} (L - s)}
        \ge \frac{9-\sqrt{57}}{4} - \frac{9-\sqrt{57}}{2} \epsilon
    \]
    when $s \le \frac{L}2$; if $s \ge \frac{L}{2}$ then Alice terminates immediately (since $s \ge \frac{L}{2} \ge \frac{\sqrt{57}-1}{8} N$), and there must've been at least $s$ corruptions, for a rate of $\ge \frac{s}{N + L} \ge \frac{L/2}{N+L} \ge \frac{9-\sqrt{57}}{2}$.
    
    Now, we consider if the adversary corrupts messages so that Bob outputs incorrectly. We may assume that the adversary corrupts fewer than $\frac{\sqrt{57}-1}{8} N$ of Bob's messages to Alice: otherwise, if the adversary corrupts $k \ge \frac{1-\epsilon}{2} L$ of Bob's messages to Alice, then if $s \ge \frac{1-\epsilon}{2} L$, Alice terminates immediately after Bob's $L$ rounds, and there must've been $\ge s$ corruptions to Bob's messages, for a total corruption rate of $\frac{\frac{1-\epsilon}2 L}{N + L} = \frac{9-\sqrt{57}}{4} - \frac{9-\sqrt{57}}{4} \epsilon$. In the other case that $s < \frac{1-\epsilon}{2} L$, it holds that $s \ge (1-\epsilon)L - k$, meaning that Alice speaks at most $M - \frac{4}{9-\sqrt{57}} s \le \frac{4}{9-\sqrt{57}} k - \frac{7+\sqrt{57}}{4}(1-2\epsilon) N$ additional rounds. This gives a corruption rate of 
    \[
        \ge \frac{k}{N + M + \frac{4}{9-\sqrt{57}} k - \frac{7+\sqrt{57}}{4}(1-2\epsilon) N}
        \ge \frac{k}{\frac{4}{9-\sqrt{57}} k + \frac{7+\sqrt{57}}{2}\epsilon N}
        \ge \frac{9-\sqrt{57}}{4}.
    \]
    Next, if the adversary corrupts $k \in [\frac{\sqrt{57}-1}{8}N, \frac{1-\epsilon}2 L)$ of Bob's messages to Alice, then $s = k$ and Alice terminates immediately at the end of Bob's $L$ rounds and has learned Bob's input correctly. Bob will also learn Alice's correct input unless at least $\frac{1-\epsilon}{2} N$ of Alice's messages to him (including the later rounds where Alice did not actually speak) are corrupted. In order for this to happen, the adversary must've corrupted at least a
    \[
        \frac{k + \frac{1-\epsilon}{2} N}{N + L}
        = \frac{(\frac{3 + \sqrt{57}}{8} - \frac{\epsilon}{2})\cdot N}{\frac{7+\sqrt{57}}{4} N}
        = \frac{9-\sqrt{57}}{4} - \frac{2}{7+\sqrt{57}} \epsilon
    \]
    fraction of total messages sent.
    
    Thus, we assume that the adversary corrupted fewer than $\frac{\sqrt{57}-1}{8} N$ of Bob's messages to Alice. There are three ways that Bob may output incorrectly. The first is that $\hat{x}_1 = x$ but Bob outputs $\hat{x}_2 \not= \hat{x}_1$. The fact that $\hat{x}_1 = x$ and the adversary corrupted $< \frac{\sqrt{57}-1}{N} < \frac{1-\epsilon}{2} L$ means that Alice sees $x' = x$, so she terminates immediately. This gives a total of $N + L$ rounds where a party speaks. In order for Bob to output $\hat{x}_2 \not= \hat{x}_1$, at least $\frac{3+\sqrt{57}}{8} N$ of Alice's last chunk of messages must be non-$\perp$ characters. This means that the adversary corrupted at least $\frac{3+\sqrt{57}}{8} N$ messages, for a corruption rate of 
    \[
        \ge \frac{\frac{3+\sqrt{57}}{8} N}{N + L}
        = \frac{9-\sqrt{57}}{4}.
    \]
    
    The second case is that $\hat{x}_1 \not= x$ but Bob outputs $\hat{x}_1$. This can only happen if Bob hears less than $\frac{3+\sqrt{57}}{8} N - t$ non-$\perp$ characters in Alice's last chunk. Suppose that the adversary corrupts $s < \frac{\sqrt{57}-1}{8} N$ of Bob's messages to Alice, so that Alice speaks $\frac{7+\sqrt{57}}{4} N - \frac{4}{9-\sqrt{57}} s$ additional rounds. Since Bob hears less than $\frac{\sqrt{57}-1}{8} N$ non-$\perp$ messages, the adversary must've corrupted $> \frac{7+\sqrt{57}}{4} N - \frac{4}{9-\sqrt{57}}s - (\frac{3+\sqrt{57}}{8} N - t) = \frac{11+\sqrt{57}}{8} N - \frac{4}{9-\sqrt{57}}s + t$ messages in Alice's second chunk. Combining with the $\ge (1-\epsilon)N - t$ corruptions the adversary must've made to Alice's first chunk of messages, and the $s$ corruptions to Bob's messages, this is a total corruption count of $> (\frac{19+\sqrt{57}}{8} - \epsilon) N - \frac{3+\sqrt{57}}{6} s$, giving a corruption rate of 
    \begin{align*}
        > \frac{(\frac{19+\sqrt{57}}{8} - \epsilon) \cdot N - \frac{3+\sqrt{57}}{6} s}{N + L + \frac{7+\sqrt{57}}{4} N - \frac{4}{9-\sqrt{57}} s}
        &= \frac{(\frac{19+\sqrt{57}}{8} - \epsilon) \cdot N - \frac{3+\sqrt{57}}{6} s}{\frac{7+\sqrt{57}}{2} N - \frac{9+\sqrt{57}}{6} s} \\
        &\ge \frac{(\frac{19+\sqrt{57}}{8} - \epsilon) N - \frac{3+\sqrt{57}}{6} \cdot \frac{\sqrt{57}-1}{8} N}{\frac{7+\sqrt{57}}{2} N - \frac{9+\sqrt{57}}{6} \cdot \frac{\sqrt{57}-1}{8} N} \\
        &= \frac{5\sqrt{57}-37}{2} - O(\epsilon) \\
        &> \frac{9-\sqrt{57}}{4}.
    \end{align*}
    
    The final case is that $\hat{x}_1 \not= x$ but Bob outputs $\hat{x}_2 \not= \hat{x}_1, x$. In this case, the adversary must've corrupted $(1-\epsilon)N - t$ messages in common between $m_A$ and $\ECC_A(\hat{x}_1)$, as well as at least $(\frac12 t - \epsilon N) + \frac{1-\epsilon}{2} \cdot (M - \frac{4}{9-\sqrt{57}} s)$ of Alice's other messages (in both the first and second chunks). This means that the adversary must've corrupted at least
    \begin{align*}
        &\frac{s + ((1-\epsilon)N - t) + (\frac12 t - \epsilon N) + \frac{1-\epsilon}{2} \cdot (M - \frac{4}{9-\sqrt{57}} s)}{N + L + M - \frac{4}{9-\sqrt{57}} s} \\
        &= \frac{(\frac{15+\sqrt{57}}{8} - \frac{23 + \sqrt{57}}{8} \epsilon) \cdot N - \frac12 t - (\frac{\sqrt{57}-3}{12} - \frac{9+\sqrt{57}}{12}\epsilon) \cdot s}{\frac{7+\sqrt{57}}{2} N - \frac{9+\sqrt{57}}{6} s} \\
        &\ge \frac{(\frac{15+\sqrt{57}}{8} - \frac{23 + \sqrt{57}}{8} \epsilon) \cdot N - \frac12 \cdot \frac{9-\sqrt{57}}{4} N - (\frac{\sqrt{57}-3}{12} - \frac{9+\sqrt{57}}{12}\epsilon) \cdot s}{\frac{7+\sqrt{57}}{2} N - \frac{9+\sqrt{57}}{6} s} \\
        &= \frac{(\frac{3+\sqrt{57}}{4} - \frac{23 + \sqrt{57}}{8}) \cdot N - (\frac{\sqrt{57}-3}{12} - \frac{9+\sqrt{57}}{12}\epsilon) \cdot s}{\frac{7+\sqrt{57}}{2} N - \frac{9+\sqrt{57}}{6} s} \\
        &\ge \frac{9-\sqrt{57}}{4} - \frac{2\sqrt{57} - 13}{4} \epsilon \\
        &\approx \frac{9-\sqrt{57}}{4} - 0.525 \epsilon.
    \end{align*}
\end{proof}

\section{Channels with Feedback} \label{sec:feedback}

Another model in which adaptive length is natural is that of interactive communication protocols with \emph{feedback}. In a scheme with feedback, the sender of a message learns the value of the message as received by the other party. One exceptional property of such schemes is that the order of speaking need not be fixed beforehand, yet the parties can still agree on a speaker for each round based on their shared knowledge of the messages received by both sides. 

The work of~\cite{EfremenkoGH16} studied the maximal noise resilience of protocols with feedback and a non-fixed order of speaking, but with fixed protocol length. They showed that for binary (and large) alphabets, there exists a protocol in this model that is resilient to a $\frac13 - \epsilon$ fraction of errors (as a fraction of the total protocol length). They also showed that $\frac13$ is an upper bound on the maximal noise resilience, assuming fixed length. 

We note that in a scheme with feedback, the parties are not only able to agree on a (non-fixed) speaker for each round, but in fact are able to coordinate a mutual (possibly early) termination based on the transcript. A coordinated early termination was not studied by~\cite{EfremenkoGH16}.

In the binary case, we show that non-fixed termination does not improve the error resilience and prove an upper bound of $\frac13$.

However, for ternary (and larger) alphabets, allowing early termination circumvents the $\frac13$ upper bound on noise resilience. In particular, there exists a message exchange protocol in the feedback model with non-fixed speaking order and termination that is resilient to a $\frac12 - \epsilon$ fraction of corruptions. This is matched by a natural upper bound of $\frac12$. We note that our protocol \emph{does not} efficiently simulate an arbitrary noiseless protocol with error resilience approaching $\frac12$, but rather only the message exchange protocol (which permits simulating an arbitrary protocol with exponential blowup in communication). The question of whether allowing early termination circumvents the $\frac13$ upper bound for ternary or larger alphabets to \emph{efficiently} simulate an arbitrary protocol (instead of just message exchange) remains open.

\subsection{Schemes with Feedback and a Ternary Alphabet}

We construct a protocol for the message exchange procedure that is resilient to a $\frac12 - \epsilon$ fraction of errors.

    \protocol{Scheme with resilience $\frac12-\epsilon$ in the Adaptive Feedback Model}{1/2-feedback}{
        Let $n = |x| = |y|$ be the length of both Alice's and Bob's inputs (if one is shorter than the other, we can pad it with $0$'s). Let $x' = x || 0^{n/\epsilon - n}$ and $y ' = y || 0^{n/\epsilon - n}$. The protocol consists of up to $n/\epsilon^2$ rounds (the parties terminate after round $n/\epsilon^2$ if they have not already terminated).
        
        \begin{itemize}
            \item The first rounds consist of Alice speaking contiguously. Her goal is to communicate $x'$ to Bob, using the symbols $0,1,\leftarrow$. To do this, both parties keep track of the currently built value of $x'$, denoted by $\hat{x}$, which is initially the empty string $\emptyset$. If $\hat{x}$ is a prefix of $x'$, Alice sends the next bit of $x'$. Otherwise, she'll send $\leftarrow$ to ask Bob to remove the last bit of $\hat{x}$. The parties update $\hat{x}$ based on the messages Bob has received from Alice, as follows: $\hat{x}$ is initially set to $\emptyset$. For any bit $0$ or $1$ received by Bob, both parties append that bit to $\hat{x}$. For any $\leftarrow$ received by Bob, both parties remove the last bit of $\hat{x}$. If $\hat{x}$ reaches length $n/\epsilon$, it becomes Bob's turn to speak.
            \item Bob sends $y'$ to Alice, one bit at a time. Both parties keep track of a string $\hat{y}$ equal to the currently built value of $y'$ based on the messages received by Alice. If $\hat{y}$ reaches length $n/\epsilon$, both parties terminate.
        \end{itemize}
}

\begin{theorem}
    Protocol~\ref{prot:1/2-feedback} is resilient to a $\frac12 - \epsilon$ fraction of error.
\end{theorem}

\begin{proof}
    Suppose that the adversary corrupts messages so that either Alice or Bob outputs incorrectly. We will show that they corrupted at least $\frac12 - \epsilon$ of all the messages. 
    
    First, note that if $p$ of a party's messages are uncorrupted and $q$ are corrupted, then the other party learns the correct value of their enlargened input (either $x'$ or $y'$) as long as $p - q = n/\epsilon$. On the other hand, the other party will learn the incorrect value only if $q - p \ge n/\epsilon - n + 1$. In particular, this means that if a party speaks for at least $k$ rounds, then at least $\frac{k-n/\epsilon}2$ rounds were corrupted. If the party speaks for exactly $k$ rounds, then the other party learns the incorrect value when at least $\frac{k+n/\epsilon - n + 1}2$ rounds are corrupted. 
    
    This means that if the adversary corrupts messages so that the parties have not terminated by round $n/\epsilon^2$, then at least $\frac{n/\epsilon^2 - 2n/\epsilon}{2}$ rounds are corrupted. In particular, this means that at least $\frac12 - \epsilon$ of the total $n/\epsilon^2$ rounds are corrupted.
    
    Otherwise, suppose that the parties terminate earlier than round $n/\epsilon^2$. Let $k_A$ be the number of rounds that Alice speaks, and let $k_B$ be the number of rounds that Bob speaks so that $k_A + k_B < n/\epsilon^2$ and $k_A, k_B \ge \frac{n}{\epsilon}$. If the adversary corrupts messages so that Alice outputs incorrectly, then they must've corrupted at least $\frac{k_A - n/\epsilon}{2} + \frac{k_B + n/\epsilon - n + 1}{2} = \frac{k_A + k_B - (n-1)}{2}$ rounds, which is a $\frac12 - \frac{n-1}{2(k_A + k_B)} \ge \frac12 - \frac\epsilon4$ fraction of the $k_A + k_B$ rounds in which the parties participate. The case where the adversary corrupts messages so that Bob outputs incorrectly is similar.
\end{proof}

\subsection{Schemes with Feedback and a Binary Alphabet} 

\begin{theorem}
    Any feedback protocol with adaptive termination that computes the identity function $f(x, y) = (x, y)$ over a feedback channel with an error rate of $\frac13$, cannot be guaranteed to succeed.
\end{theorem}

\begin{proof}
    Let Alice have possibilities $x_1, x_2, x_3$ and Bob have $y_1, y_2, y_3$. At every step in the protocol, regardless of who speaks, the adversary will choose to send the majority bit among the 3 possible inputs. Thus, the transmission will be the same regardless of what inputs each person actually has. 
    
    It remains to show that at any point $N$ in the protocol, for some choice of input that one party has, there are two options for the other party that both have resulted in less than $\frac13N$ corruption so far. For $i\in \{1,2,3\}$, let $X_i$ (resp. $Y_i$) denote the number of times Alice's (resp. Bob's) message was corrupted in the case that she had $x_i$ (resp. $y_i$) so far.
    
    Since only one of the six corruptions can occur in each round, we have that 
    \[
        X_1+X_2+X_3+Y_1+Y_2+Y_3\leq N.
    \]
    Then, without loss of generality, we can let 
    \[
        (X_1+Y_1)\leq (X_2+Y_2)\leq (X_3+Y_3)
    \] 
    which implies that \[
        X_1+X_2+Y_1+Y_2\leq \frac23N.
    \] 

    Without loss of generality, let $X_1\leq X_2$ and $Y_1 \leq Y_2$. The previous equation shows that one of $X_1+Y_2$ and $Y_1+X_2$ is at most $\frac13N$. In the former case, Alice corruption was within the budget regardless of whether Bob had $y_1$ or $y_2$ when she has $x_1$ (because $X_1+Y_1 \leq \frac13N$ and $X_1+Y_2 \leq \frac13N$). In the latter case, Bob can't distinguish between $x_1$ and $x_2$ when Alice has $y_1$.
\end{proof}

\bibliographystyle{alpha}
\bibliography{refs}

\newcommand{\etalchar}[1]{$^{#1}$}
\begin{thebibliography}{DHM{\etalchar{+}}15}

\bibitem[AGS16]{AgrawalGS16}
Shweta Agrawal, Ran Gelles, and Amit Sahai.
\newblock Adaptive protocols for interactive communication.
\newblock In {\em {IEEE} International Symposium on Information Theory, {ISIT}
  2016, Barcelona, Spain, July 10-15, 2016}, pages 595--599. {IEEE}, 2016.

\bibitem[BE14]{BravermanE14}
M.~Braverman and K.~Efremenko.
\newblock List and unique coding for interactive communication in the presence
  of adversarial noise.
\newblock In {\em 2014 IEEE 55th Annual Symposium on Foundations of Computer
  Science (FOCS)}, pages 236--245, Los Alamitos, CA, USA, oct 2014. IEEE
  Computer Society.

\bibitem[BK12]{BrakerskiK12}
Zvika Brakerski and Yael~Tauman Kalai.
\newblock Efficient interactive coding against adversarial noise.
\newblock In {\em 2012 IEEE 53rd Annual Symposium on Foundations of Computer
  Science}, pages 160--166, 2012.

\bibitem[BN13]{BrakerskiN13}
Zvika Brakerski and Moni Naor.
\newblock Fast algorithms for interactive coding.
\newblock In {\em Proceedings of the Twenty-Fourth Annual ACM-SIAM Symposium on
  Discrete Algorithms}, SODA '13, page 443–456, USA, 2013. Society for
  Industrial and Applied Mathematics.

\bibitem[BR11]{BravermanR11}
Mark Braverman and Anup Rao.
\newblock Towards coding for maximum errors in interactive communication.
\newblock In {\em Proceedings of the Forty-Third Annual ACM Symposium on Theory
  of Computing}, STOC '11, page 159–166, New York, NY, USA, 2011. Association
  for Computing Machinery.

\bibitem[Bra12]{Braverman12}
Mark Braverman.
\newblock Towards deterministic tree code constructions.
\newblock In {\em Proceedings of the 3rd Innovations in Theoretical Computer
  Science Conference}, ITCS '12, page 161–167, New York, NY, USA, 2012.
  Association for Computing Machinery.

\bibitem[DHM{\etalchar{+}}15]{DaniHMSY15}
Varsha Dani, Thomas~P. Hayes, Mahnush Movahedi, Jared Saia, and Maxwell Young.
\newblock Interactive communication with unknown noise rate, 2015.

\bibitem[EGH16]{EfremenkoGH16}
Klim Efremenko, Ran Gelles, and Bernhard Haeupler.
\newblock Maximal noise in interactive communication over erasure channels and
  channels with feedback.
\newblock {\em {IEEE} Trans. Inf. Theory}, 62(8):4575--4588, 2016.

\bibitem[EKPS23]{EfremenkoKPS23}
Klim Efremenko, Gillat Kol, Dmitry Paramonov, and Raghuvansh~R Saxena.
\newblock Protecting single-hop radio networks from message drops.
\newblock In {\em 50th International Colloquium on Automata, Languages, and
  Programming (ICALP 2023)}. Schloss Dagstuhl-Leibniz-Zentrum f{\"u}r
  Informatik, 2023.

\bibitem[EKS20a]{EfremenkoKS20b}
Klim Efremenko, Gillat Kol, and Raghuvansh~R. Saxena.
\newblock Binary interactive error resilience beyond ${{}^{1}}\!/\!_{8}$ (or
  why $({{}^{1}}\!/\!_{2})^{3} > {{}^{1}}\!/\!_{8})$.
\newblock In {\em 2020 IEEE 61st Annual Symposium on Foundations of Computer
  Science (FOCS)}, pages 470--481, 2020.

\bibitem[EKS20b]{EfremenkoKS20}
Klim Efremenko, Gillat Kol, and Raghuvansh~R. Saxena.
\newblock Interactive error resilience beyond 2/7.
\newblock In Konstantin Makarychev, Yury Makarychev, Madhur Tulsiani, Gautam
  Kamath, and Julia Chuzhoy, editors, {\em Proccedings of the 52nd Annual {ACM}
  {SIGACT} Symposium on Theory of Computing, {STOC} 2020, Chicago, IL, USA,
  June 22-26, 2020}, pages 565--578. {ACM}, 2020.

\bibitem[EKS21]{EfremenkoKS21a}
Klim Efremenko, Gillat Kol, and Raghuvansh Saxena.
\newblock Optimal error resilience of adaptive message exchange.
\newblock {\em Electron. Colloquium Comput. Complex.}, 28:60, 2021.

\bibitem[Gel17]{Gelles17}
Ran Gelles.
\newblock Coding for interactive communication: A survey.
\newblock {\em Foundations and Trends® in Theoretical Computer Science},
  13:1--161, 01 2017.

\bibitem[GH13]{GhaffariH13}
Mohsen Ghaffari and Bernhard Haeupler.
\newblock Optimal error rates for interactive coding ii: Efficiency and list
  decoding.
\newblock {\em Proceedings - Annual IEEE Symposium on Foundations of Computer
  Science, FOCS}, 12 2013.

\bibitem[GH17]{GellesH17}
Ran Gelles and Bernhard Haeupler.
\newblock Capacity of interactive communication over erasure channels and
  channels with feedback.
\newblock {\em SIAM Journal on Computing}, 46:1449--1472, 01 2017.

\bibitem[GHK{\etalchar{+}}16]{GellesHKRW16}
Ran Gelles, Bernhard Haeupler, Gillat Kol, Noga Ron-Zewi, and Avi Wigderson.
\newblock {\em Towards Optimal Deterministic Coding for Interactive
  Communication}, pages 1922--1936.
\newblock 2016.

\bibitem[GHS14]{GhaffariHS14}
Mohsen Ghaffari, Bernhard Haeupler, and Madhu Sudan.
\newblock Optimal error rates for interactive coding {I:} adaptivity and other
  settings.
\newblock In David~B. Shmoys, editor, {\em Symposium on Theory of Computing,
  {STOC} 2014, New York, NY, USA, May 31 - June 03, 2014}, pages 794--803.
  {ACM}, 2014.

\bibitem[GI18]{GellesI18}
Ran Gelles and Siddharth Iyer.
\newblock Interactive coding resilient to an unknown number of erasures.
\newblock {\em arXiv preprint arXiv:1811.02527}, 2018.

\bibitem[GZ22]{GuptaZ22a}
Meghal Gupta and Rachel~Yun Zhang.
\newblock {The Optimal Error Resilience of Interactive Communication Over
  Binary Channels}.
\newblock In {\em Symposium on Theory of Computing, {STOC} 2012, New York, NY,
  USA, June 20 - June 24, 2022}, STOC '22. {ACM}, 2022.

\bibitem[GZ23]{GuptaZ23}
Meghal Gupta and Rachel~Yun Zhang.
\newblock Efficient interactive coding achieving optimal error resilience over
  the binary channel.
\newblock In {\em Proceedings of the 55th Annual ACM Symposium on Theory of
  Computing}, pages 1449--1462, 2023.

\bibitem[Hae14]{Haeupler14}
Bernhard Haeupler.
\newblock Interactive channel capacity revisited.
\newblock In {\em 55th {IEEE} Annual Symposium on Foundations of Computer
  Science, {FOCS} 2014, Philadelphia, PA, USA, October 18-21, 2014}, pages
  226--235, 2014.

\bibitem[Ham50]{Hamming50}
R.~W. Hamming.
\newblock Error detecting and error correcting codes.
\newblock {\em The Bell System Technical Journal}, 29(2):147--160, 1950.

\bibitem[Sch92]{Schulman92}
L.J. Schulman.
\newblock Communication on noisy channels: a coding theorem for computation.
\newblock In {\em Proceedings., 33rd Annual Symposium on Foundations of
  Computer Science}, pages 724--733, 1992.

\bibitem[Sch93]{Schulman93}
Leonard~J. Schulman.
\newblock Deterministic coding for interactive communication.
\newblock In {\em Proceedings of the Twenty-Fifth Annual ACM Symposium on
  Theory of Computing}, STOC '93, page 747–756, New York, NY, USA, 1993.
  Association for Computing Machinery.

\bibitem[Sch96]{Schulman96}
Leonard~J Schulman.
\newblock Coding for interactive communication.
\newblock {\em IEEE Transactions on Information Theory}, 42(6):1745–1756,
  1996.

\bibitem[Sha48]{Shannon48}
C.~E. Shannon.
\newblock A mathematical theory of communication.
\newblock {\em The Bell System Technical Journal}, 27(3):379--423, 1948.

\end{thebibliography}

\end{document}